\documentclass[]{emulateapj}

\usepackage{booktabs}
\usepackage{amsmath}
\usepackage{multirow}
\usepackage{booktabs,array,dcolumn}
\usepackage{tabularx}
\usepackage[export]{adjustbox}
\newcolumntype{d}{D{.}{.}{2.3}}
\newcolumntype{C}{>{\centering}p}

\def\hi{{\mbox{\sc Hi}}}
\def\galaxy{NGC~1052--DF2}
\def\msun{$\rm M_{\odot}$}
\def\lsun{$\rm L_{\odot}$}
\def\mstar{$M_{\star}$}

\shorttitle{NGC1052-DF2}
\shortauthors{Sardone {\em et al.}}

\begin{document}

\title{Constraints on the \hi \, Mass for \galaxy{}}

\author{Amy Sardone\textsuperscript{1,2}} 
\author{D.J. Pisano\textsuperscript{1,2,3}} 
\author{Sarah Burke-Spolaor\textsuperscript{1,2}} 
\author{Joshua L. Mascoop\textsuperscript{1,2}} 
\author{Nihan Pol\textsuperscript{1,2}} 
\affil{\textsuperscript{1}{Department of Physics and Astronomy, 
West Virginia University, Morgantown, WV 26506}}
\affil{\textsuperscript{2}Gravitational Wave and Cosmology Center, Chestnut Ridge Research Building, Morgantown, WV 26505 USA}
\affil{\textsuperscript{3}Adjunct Astronomer at Green Bank Observatory}
\email{asardone@mix.wvu.edu}

\begin{abstract}
We report deep, single-dish 21 cm observations of \galaxy{}, taken with the Green Bank Telescope. \galaxy{}, proposed to be lacking in dark matter, is currently classified as an ultra-diffuse galaxy in the NGC~1052 group. We do not detect the galaxy, and derive an upper limit on the \hi \, mass. The galaxy is extremely gas-poor, and we find that a $3\sigma \, M_{\hi}$ detection at a distance of 19 Mpc and using a line width of 3.2 km $\rm s^{-1}$ would have an upper limit of $M_{\hi,lim} < 5.5 \times 10^5$ \msun. At this mass limit, the gas fraction of neutral gas mass to stellar mass is extremely low, at $M_{\hi}$/\mstar $\, < \, 0.0027$. This extremely low gas fraction, comparable to Galactic dwarf spheroidals and gas-poor dwarf ellipticals, implies that either the galaxy is within the virial radius of NGC~1052, where its gas has been stripped due to its proximity to the central galaxy, or that \galaxy{} is at distance large enough to inhibit detection of its gas. We also estimated the upper limit of the \hi \, mass of \galaxy{} resided at 13 Mpc. This would give an \hi \, mass of $M_{\hi,lim} < 2.5 \times 10^5$ \msun, and an \hi \, gas fraction of $M_{\hi}$/\mstar $\, < \, 0.0012$, becoming even more extreme. While the dark matter fraction would be less extreme at this distance, the neutral gas fraction would be unprecedented for an object in a low density environment. 
\end{abstract}


\section{Introduction}



The extremely low surface brightness galaxy \galaxy{} was discovered by \cite{2000A&AS..145..415K}, who labeled it a dwarf galaxy candidate. \cite{no_dm_galaxy} measured the total mass of the galaxy by measuring the radial velocities of ten luminous globular clusters. Using the inferred velocity dispersion, \cite{no_dm_galaxy} determined the total mass within a 7.6 kpc radius to be less than $3.4 \times 10^8 \, \rm M_\odot$. These globular clusters trace the mass profile of \galaxy{} out to radii nearly as large as the virial radius of the galaxy ($\sim 10$ kpc). The dark matter halo mass can be estimated using the dark matter halo mass/stellar mass ratio $M_{\rm halo}/$\mstar{}, where the expected $M_{\rm halo}/$\mstar \, ratio for low mass galaxies like \galaxy{} is greater than 30 (\cite{2010ApJ...710..903M,2013ApJ...770...57B}). Comparing the estimated total mass with the derived stellar mass of the galaxy, which \cite{no_dm_galaxy} determined to be \mstar{} $ \approx 2 \times 10^8 \,$\msun{}, they obtain a $M_{\rm halo}/$\mstar{} of order one. Thus, they propose that the galaxy is deficient in dark matter. 

If \galaxy{} is truly a galaxy lacking dark matter, the question of how dark matter is separated from baryonic matter remains. \cite{2006ApJ...648L.109C} showed that dark matter can be dissociated from galaxies if dark matter is bound to baryons through nothing but gravity. However, until now, previous attempts have not been fruitful in observing a galaxy without dark matter \citep{2003Sci...301.1696R, 2014MNRAS.440.1634P}. 

Recently, \cite{2018arXiv180404139L} suggested a lack of robustness in the method used by \cite{no_dm_galaxy} to obtain the mass to light ratio, M/L, by using the globular clusters in \galaxy{}. They show that similar methods applied to the well-studied Fornax dSph would give wildly different dark matter halo mass estimates with large scatter in the velocity dispersion at the 95\% confidence level.

\cite{2018arXiv180610141T} proposed that many of the unusual features of \galaxy{} may be explained if the galaxy, which \cite{no_dm_galaxy} estimates a distance of 19 Mpc, was brought to a distance of 13 Mpc, making it a typical low surface brightness galaxy without the anomalies described by \cite{no_dm_galaxy}. \cite{2018ApJ...864L..18V} address this distance concern by analyzing the color-magnitude diagram (CMD) of \galaxy{} and arrive at a distance consistent with the 19 Mpc estimate. They provide an additional distance estimate by applying a method free of calibration uncertainties and again arrive at the same 19 Mpc distance estimate. \cite{2018RNAAS...2c.146B} performed an independent analysis of the distance with similar conclusions of $\rm D = 20.4 \pm 2.0 \, Mpc$. In this paper, we provide the 21 cm neutral hydrogen (\hi{}) mass upper limit calculations using the 19 Mpc distance estimate.

Most recently, \cite{2019MNRAS.482L..99C} found upper limits on the \hi \, mass to be $M_{\hi,lim} < 3.15 \times 10^6$ \msun with 20 km/s resolution. Our observation with a single dish allowed us to go deeper, probing the extreme nature of this source, obtaining a more constrained upper limit.  

This paper proceeds as follows. In Section 2 we describe the parameters of our observations using the GBT. In Section 3 we calculate the upper limits and describe our analysis of the data. In Section 4 we present our results, and we conclude in Section 5 with discussion of the significance of these results for \galaxy.

\begin{figure*}[t]
\centering
        \vspace{-1cm}
        \includegraphics[width = 0.8\textwidth]{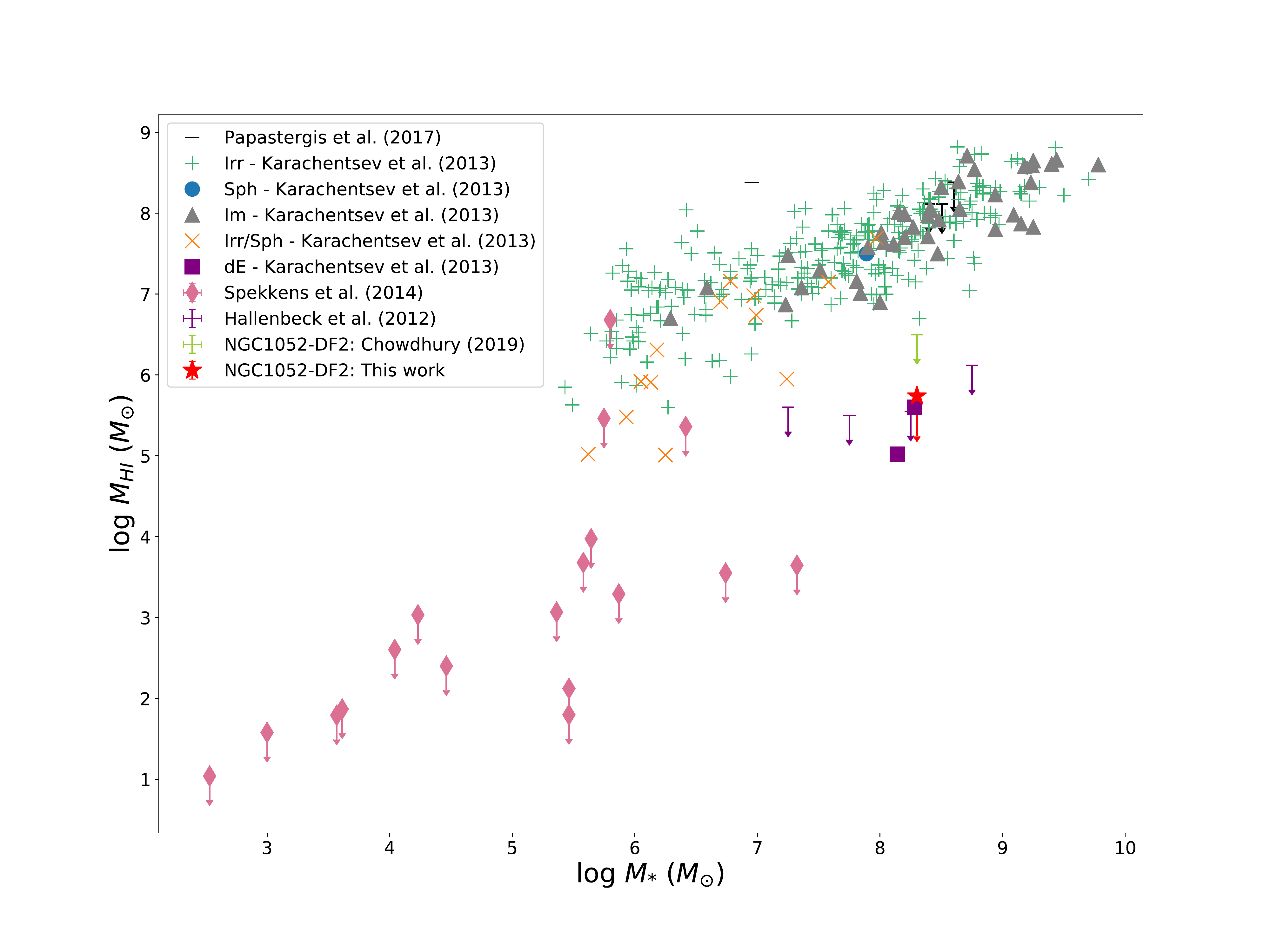}
        
        \includegraphics[width = 0.8\textwidth]{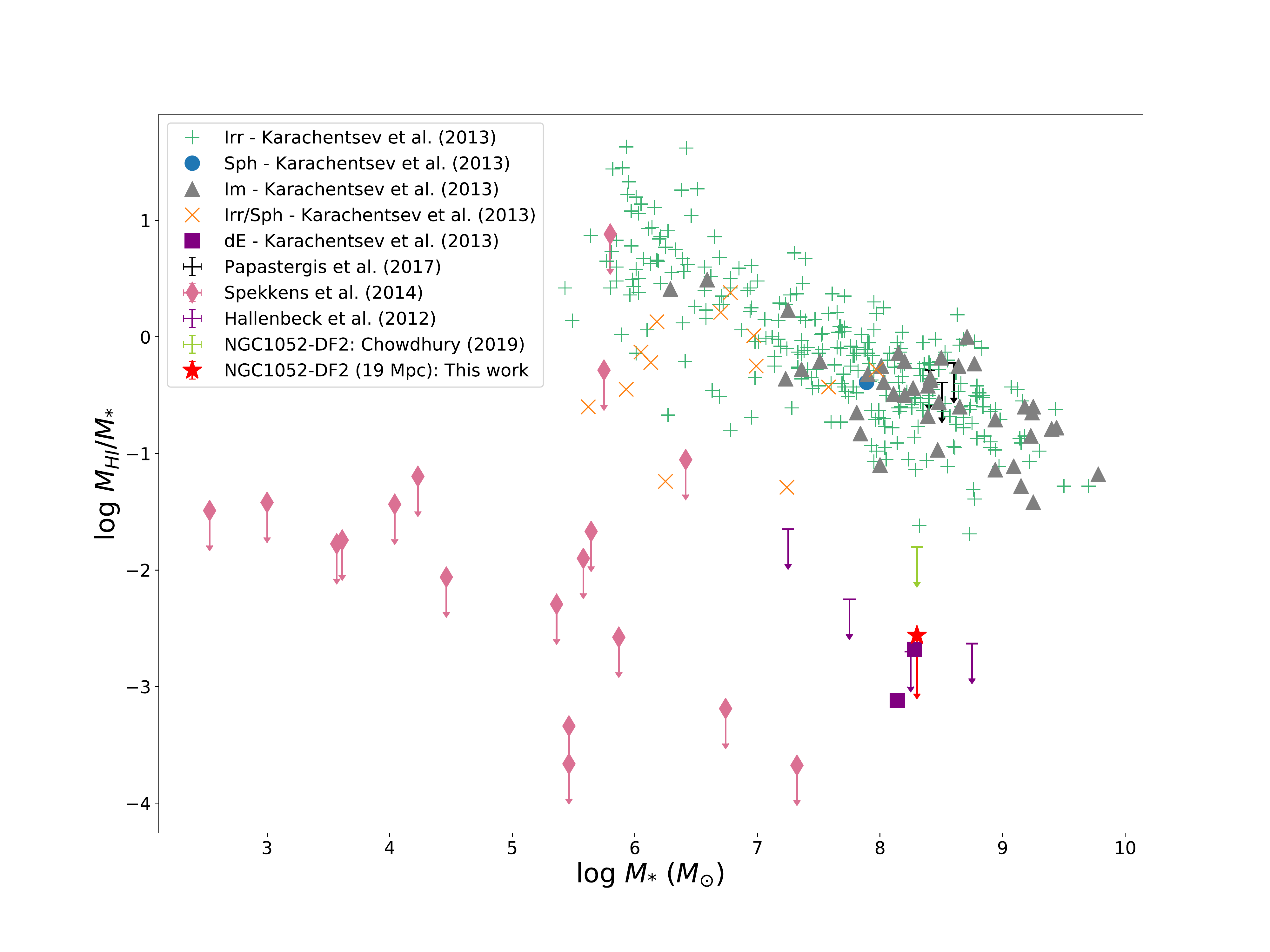}
        \caption{{\it Top}: Stellar mass-\hi \, mass relation for dwarf galaxies. The black upper limits from \cite{2017A&A...601L..10P} represent isolated UDGs at 40-80 Mpc. The pink diamonds represent the upper limits for Galactic dSphs and Local Group dSphs from \cite{2014ApJ...795L...5S} (nearly all are $<$ 1 Mpc). Crosses, circles, triangles, Xs, and squares represent the various morphologies of dwarf galaxies within 11 Mpc \citep{2013AJ....145..101K}. Purple upper limits from \cite{2012AJ....144...87H} represent the dwarf ellipticals and dwarf lenticulars (dE, dS0) in the Virgo cluster (D$\sim 17$ Mpc). The previous upper limit on the HI mass of NGC1052-DF2 by \cite{2019MNRAS.482L..99C} is shown in yellow-green. An updated HI mass upper limit for NGC1052-DF2 from this work is shown for the distance of 19 Mpc in red. {\it Bottom}: Relationship between the stellar mass and the HI gas fraction for the sample in the figure above. Symbols remain the same. Apart from the extremely nearby ($<$ 1 Mpc) dSphs from \cite{2014ApJ...795L...5S}, the extreme nature of the gas fraction of NGC1052-DF2 becomes clear, as it is a galaxy in a low-density environment with a comparable neutral gas fraction to those in a high-density cluster environment.}
\end{figure*}

\clearpage

\section{Observations}

\begin{table*}[t]
\caption{Properties of \galaxy{}}
\small
\centering
    \begin{tabular}{cccccccc}
    \hline
    \hline
     \addlinespace[0.2cm]
      $\Delta v$ \footnotemark[1] & $\sigma_{rms}$ \footnotemark[2] & $S_{HI,lim}$ \footnotemark[3] & M$_{\hi}^{lim}\, [19 \rm \,Mpc]$ & $M_{\hi}^{lim}/$\mstar{} & $M_{\hi}^{lim}/L_V$ & $M_{\hi}^{lim}/M_{dyn}$ \\
         (km/s) & (mJy/beam) & ($\rm Jy \, km \, s^{-1}$) &(\msun) & & (\msun/\lsun)\\
     \addlinespace[0.2cm]
     \hline
     \addlinespace[0.1cm]
    3.2 & 0.673 & 0.006 & $< 5.5 \times 10^5$ & $< 0.0027$ & $< 0.005$ & $< 0.0016$ \\
    \addlinespace[0.1cm]
    \hline
    \hline
    \end{tabular}
\label{dwarf.tab}
\footnotetext[1]{Velocity resolution}
\footnotetext[2]{Measured rms noise}
\footnotetext[3]{Integrated flux limit}
\end{table*}
\normalsize

We searched for 21 cm (1.42 GHz) \hi{} line emission from \galaxy{} using the Robert C. Byrd Green Bank Telescope (GBT) in August 2018 (project GBT18A-508). We used the L-band (1.15-1.73 GHz) receiver with the VErsatile GBT Astronomical Spectrometer (VEGAS) backend in spectral line mode. At these frequencies, the FWHM beamwidth is $9.1'$. 

Using globular clusters in \galaxy{}, van Dokkum et al. (2018) showed that the intrinsic velocity dispersion measured, was $\sigma_v = 3.2_{-3.2}^{+5.5}\, \rm km \, s^{-1}$. Thus, we would expect the rotational velocity of NGC 1052–DF2 to be of the same order of magnitude. This requires a velocity resolution smaller than $\sigma_v$ in order to measure an accurate \hi \, line profile. As a result, we aimed for a velocity resolution of $\Delta v < 1 \, \rm km \, s^{-1}$ in the source rest frame.

To achieve a $1\sigma$ sensitivity of $\sigma_{rms} < 1 \, \rm mJy$ with the observing setup described above, we tracked \galaxy{} for a total observing time of 4 hours and 15 minutes with the GBT. We observed over a bandwidth of 100 MHz and 131072 channels, resulting in the native resolution of 0.76 kHz, or 0.16 km/s. We searched over the bandwidth for \hi \, emission at a wide range of velocities (0-11000 km/s) with a focus on the range around 1803 km/s, corresponding to an optical redshift of z $\sim 0.006$. We reduced the data using $\it getfs$ in GBTIDL and averaged and baselined each spectrum obtained. We followed this procedure by smoothing the averaged data to multiple velocity resolutions. These are displayed in Figure 1 where there is no obvious \hi \, signal detected.

We performed a search with the NASA/IPAC Extragalactic Database (NED\footnotemark[2]) using a $9'$ search radius around \galaxy{}, revealing no other likely sources of contamination at redshifts we can detect within the beam radius.
\footnotetext[2]{http://ned.ipac.caltech.edu/}

\section{Results}

We calculated our \hi \, flux upper limit using
\begin{equation}
    S_{HI,lim} = 3\, \sigma_{rms} \, \sqrt{W \, dv} \, 
\end{equation}

\noindent where $\sigma_{rms}$ is the measured noise in Jy/beam, W is the expected line width in $\rm km \, s^{-1}$, and $dv$ is the velocity resolution in $\rm km \, s^{-1}$. The flux upper limit is in units of Jy km $\rm s^{-1}$. 

The \hi \, mass of a source can be calculated using

\begin{equation}
M_{\hi} = 2.36 \times 10^5 \, D^2 \int_{0}^\infty S(v) dv \, \rm M_{\odot} ,
\end{equation}

\noindent where $\it D$ is the distance to the source in Mpc and $\int_{0}^\infty S(v) dv$ is the integrated \hi \, flux over the source with units of Jy km $\rm s^{-1}$. 

We determined the upper limit of detectable \hi \, with the requirement of a $3\sigma$ detection using 

\begin{equation}
M_{\hi,lim} = 2.36 \times 10^5 \, D^2 \, S_{HI,lim} \, \rm M_{\odot} .
\end{equation}

We chose to use a line width W, consistent with that of the line widths from kinematic measurements of the globular cluster system within \galaxy{} in \cite{no_dm_galaxy} ($W = \sigma_v = 3.2_{-3.2}^{+5.5}\, \rm km \, s^{-1}$), and smoothed our 0.16 km/s native resolution data to $\Delta v$ = 1, 3.2, 5, and 8 km/s, all within the range of errors in $\sigma_v$. Mass calculations in this paper are made using the 3.2 km/s resolution data, with the intent to increase our signal-to-noise ratio. We also present mass upper limits using line widths of $10.5 \, \rm km \, s^{-1}$ and $20 \, \rm km \, s^{-1}$ given in \cite{2018arXiv181207345E} and \cite{2019MNRAS.482L..99C}, respectively. Using the line width of $10.5 \, \rm km \, s^{-1}$, our calculated upper limit would become $M_{\hi,lim} < 9.9 \times 10^5$ \msun. A direct comparison to the limit found by \cite{2019MNRAS.482L..99C} would give us a limit of $M_{\hi,lim} < 1.6 \times 10^6$ \msun, a factor of $> 2$ better. We searched throughout our 100 MHz bandwidth at each smoothed resolution and did not detect a signal at any velocity. The noise in each spectra goes down as expected, by $\sim\sqrt{N}$, where $N$ is the number of channels being smoothed. 

For comparison, we include ratios of the \hi \, mass upper limit by the stellar mass \mstar, the total V-band luminosity $L_V$, and the dynamical mass $M_{dyn}$ in Table 1.

\begin{figure*}[t]
        \centering
        \includegraphics[width = \textwidth]{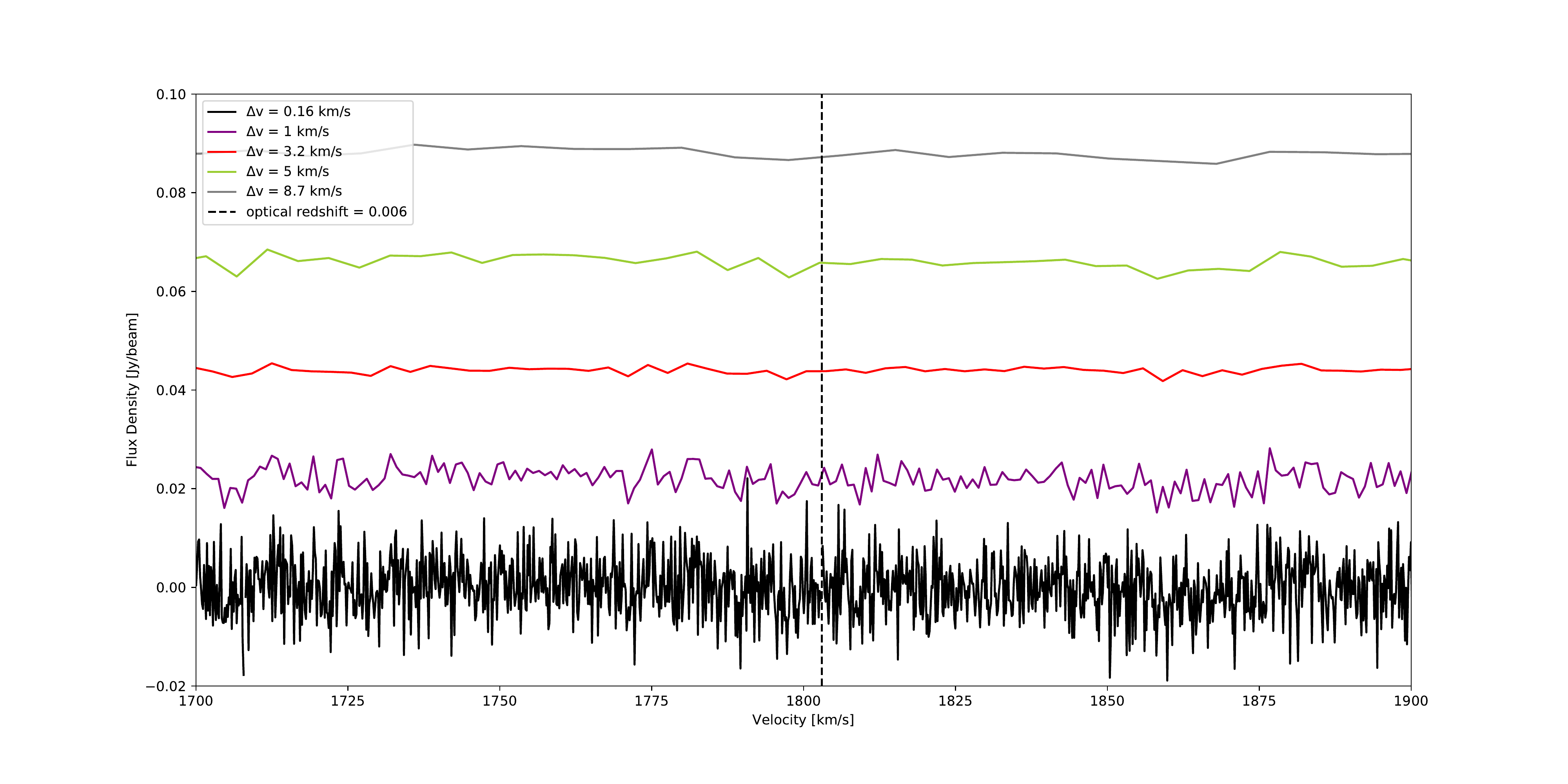}
        \caption{The averaged 4-hour data set showing 100 km/s on either side of the proposed velocity (1803 km/s) of \galaxy{}. The smoothed data, each offset by 22 mJy, is shown in various colors above the native resolution data in black. Note the 3.2 km/s velocity resolution in red, the velocity resolution of the globular cluster system found in \cite{no_dm_galaxy}, which we used for our calculations and should have produced the greatest signal-to-noise.}
        \label{non-detection}
        \vspace{0.5cm}
\end{figure*}

\begin{figure}[]
        \vspace{0.5cm}
        \includegraphics[width =0.5\textwidth, right]{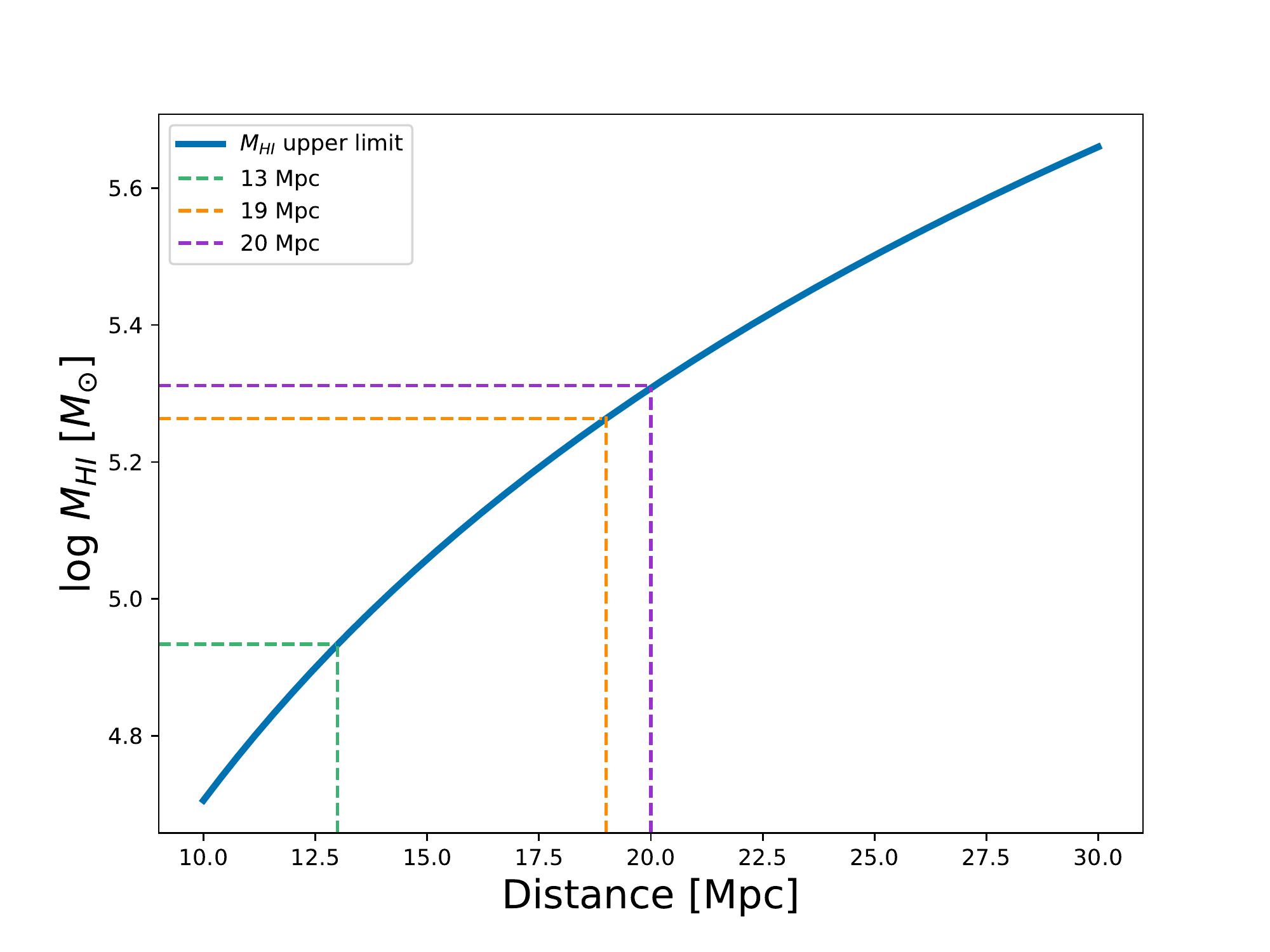}
        \caption{The blue line is our upper limit of the \hi \, mass as a function of distance, as calculated by equation (2). The sea green dashed line marks the 13 Mpc distance as proposed by \cite{2018arXiv180610141T}, the orange dashed 19 Mpc by \cite{no_dm_galaxy}, and the purple dashed 22 Mpc by \cite{2018RNAAS...2c.146B}.}
        \label{distance}
\end{figure}

\section{Discussion and Conclusions}

We have included a figure of our $M_{\hi}^{lim}$ as a function of distance (Fig. 2), encompassing the three proposed distances mentioned in this paper \citep{2018arXiv180610141T, no_dm_galaxy, 2018RNAAS...2c.146B}. All prove to be very gas-poor, with a factor of $\sim 2$ difference in \hi \, mass between the three distance estimates.

We calculate the upper limit on the $\rm M_{\hi}$ for the distance of 19 Mpc (as proposed by \cite{no_dm_galaxy}). We also calculate our integrated
flux limit $S_{HI,lim}$ using a $3\sigma$ detection limit, the \hi \, gas fraction, $M_{HI}/$\mstar{}, the \hi \, mass to V-band luminosity 
$M_{\hi}^{lim}/L_V$, and the \hi \, mass to dynamical mass ratio $M_{\hi}^{lim}/M_{dyn}$, where values for \mstar{}$\approx 2 \times 10^8$ \msun{},
$L_V=1.1 \times 10^8$ \lsun, and $M_{dyn} < 3.4 \times 10^{8}$ \msun \, are all taken from \cite{no_dm_galaxy}. All of these ratios are below 
$1\%$, demonstrating the insignificance of the amount of neutral, atomic hydrogen in this galaxy. This new upper limit would bring the gas fraction ($M_{\hi}^{lim}/$\mstar{} $< 0.0027$) down to that of the population of gas-poor dwarf ellipticals \citep{2012AJ....144...87H}, as can be seen in Fig. 1. This limit demonstrates the highly gas-deficient nature of this galaxy. 

Previous efforts have been made to detect neutral hydrogen in very low mass galaxies around the Milky Way using the Green Bank Telescope (GBT). These Galactic dwarf spheroidals (dSphs) have $5\sigma$ upper limits of $M_{HI} < 10^4$ \msun \citep{2014ApJ...795L...5S}, while neutral hydrogen detections have been made in other dSphs at comparable distances with the Parkes radio telescope \citep{2005A&A...444..133T}. Our \hi \, mass to light ratio $M_{\hi}^{lim}/L_V$ is of a similar value to that of the dwarf spheroidal galaxies associated with the Milky Way and the Local Group \citep{2014ApJ...795L...5S}. However, our \hi \, mass to dynamical mass ratio $M_{\hi}^{lim}/M_{dyn}$ is higher than that of those same Galactic dSphs by $\sim 2$ orders of magnitude, but is on par with the Local Volume dwarfs, with a distinction between these two groups being within (Galactic dSphs) or beyond (Local Volume dwarfs) the virial radius of the Milky Way. 

The amount of gas found in a galaxy is greatly connected to its environment. An ultra-diffuse galaxy (UDG) in isolation should have a neutral gas mass of $10^7 < M_{\hi} < 10^9$ \msun \citep{2017MNRAS.467.3751B, 2017A&A...601L..10P}. In groups, similar amounts of \hi \, mass have been found in UDGs \citep{2017ApJ...836..191T, 2018ApJ...855...28S}. There is an extreme lack of neutral gas in \galaxy{} as compared to other UDGs with \hi \, measurements. 

We have considered the possibility that this source is an old tidal dwarf galaxy (TDG), collisional debris from a previous merger. These old TDGs should show both a lack of dark matter, and an unusually high metallicity for their mass, with large gas depletion time-scales \citep{2001A&A...378...51B, 2000ApJ...542..137H, 2007A&A...475..187D, 2014ApJ...782...35S}. Given the less-than-solar metallicity and gas deficient nature of \galaxy{}, we do not consider this to be a likely origin.

Our \hi \, mass upper limit, however, is consistent with the upper limits for dwarf ellipticals in the Virgo cluster found by \cite{2003ApJ...591..167C}, who reported \hi \, mass upper limits as low as $5 \times 10^5$ \msun. The gas fraction upper limit we found is also consistent with the gas fractions from dwarf ellipticals found by \cite{2012AJ....144...87H}. These similarities provide further support for \galaxy{} as a dwarf elliptical.

One likely scenario for the mechanism of gas removal in \galaxy{} is through gas stripping as a result of its proximity to NGC~1052 ($\sim 80 \, \rm kpc$ in projection). The location of the source residing within the central galaxy's virial radius is an important factor in the amount of \hi \, found in a satellite (\cite{2009ApJ...696..385G,2014ApJ...795L...5S}). Because of the extended and loosely bound nature of \hi \, in galaxies, it is more likely to be stripped from its galaxy than the stars (\cite{2006PASP..118..517B,2017ApJ...844...48P}). The lack of \hi \, we find could be indicative of \galaxy{} residing within the virial radius of NGC~1052. It is possible that the \hi \, in \galaxy{} was not detected due to the source residing at some greater distance than NGC~1052. In this case, the gas removal mechanism could be through bursts of star formation or through gas expulsion \citep{2014MNRAS.445..581H}. However, finding an isolated galaxy without \hi \, would  be an unusual scenario and would require further explanation for its gas removal. The upper limit on the gas fraction $M_{HI}/$\mstar{} and the upper limit on the ratio of \hi \, mass to dynamical mass $M_{\hi}^{lim}/M_{dyn}$ could be consistent with either environmental scenarios of stripped gas by proximity to a larger galaxy or of a field galaxy with gas loss over time. While one scenario constrains the distance of \galaxy{}, the other would prove to be an atypical finding of a galaxy without neutral gas when living in isolation. If there is any neutral gas present in \galaxy{}, the insignificant amount would contribute extremely little to the baryonic mass of the galaxy. 


We found the upper limit of \hi \, mass in \galaxy{} to be $M_{\hi,lim} < 5.5 \times 10^5$ \msun with a gas fraction of neutral gas to stellar mass of $M_{\hi}$/\mstar $\, < \, 0.0027$. Such an extreme lack of neutral gas in this galaxy is consistent with known gas-poor dwarf ellipticals, dwarf spheroidals, and tidal dwarfs. Further inspection is needed to constrain the origin and morphology of this source.

\acknowledgments

We would like to thank the referee for their feedback which helped to improve the quality of this manuscript. This research was partially supported by NSF CAREER grant AST-1149491. This research made use of the NASA/IPAC Extragalactic Database (NED). SBS was supported by NSF EPSCoR award number 1458952. We thank West Virginia University for their continued support of the operations of the GBT. The GBT is operated by the Green Bank Observatory. We thank Kristine Spekkens for providing expertise on the subject of neutral gas in UDGs.

\section{ORCID iD}
{https://orcid.org/0000-0002-5783-145X}

\bibliography{bib-new.bib}
\bibliographystyle{aasjournal}
\end{document}